%% file: main.tex
\documentclass{article}

\usepackage{style/arxiv}
\usepackage[utf8]{inputenc} 
\usepackage{hyperref}       
\usepackage{url}            
\usepackage{booktabs}       
\usepackage{amsfonts}       
\usepackage{nicefrac}       
\usepackage{microtype}      
\usepackage{lipsum}		
\usepackage{graphicx}
\usepackage{natbib}
\usepackage{doi}
\usepackage{float}
\usepackage{makecell} 
\usepackage[frozencache,cachedir=.]{minted}
\usepackage{multirow}
\usepackage{caption}
\usepackage{array}

\newcolumntype{C}[1]{>{\centering\arraybackslash}m{#1}}

\newcommand{\Section}[1]{\section{#1}}
\newcommand{\Subsection}[1]{\subsection{#1}}
\newcommand{\Subsubsection}[1]{\subsubsection{#1}}
\newcommand{\Subsubsubsection}[1]{\paragraph{#1}}

\title{LLMSecCode: Evaluating Large Language Models for Secure Coding}

\title{LLMSecCode: Evaluating Large Language Models for Secure Coding}

\author{{Anton Rydén} \\
	Department of Computer Science and Engineering \\
	Chalmers University of Technology \\
	Gothenburg 41296, Sweden \\
	\texttt{rydena@chalmers.se} \\
	\And
	{Erik Näslund} \\
	Department of Computer Science and Engineering \\
	Chalmers University of Technology \\
	Gothenburg 41296, Sweden \\
	\texttt{erikna@chalmers.se} \\
	\And
	{Elad Michael Schiller } \\
	Department of Computer Science and Engineering \\
	Chalmers University of Technology \\
	Gothenburg 41296, Sweden \\
	\texttt{elad@chalmers.se} \\
	\And
	{Magnus Almgren } \\
	Department of Computer Science and Engineering \\
	Chalmers University of Technology \\
	Gothenburg 41296, Sweden \\
	\texttt{magnus.almgren@chalmers.se} \\
}

\renewcommand{\shorttitle}{LLMSecCode: Evaluating Large Language Models for Secure Coding}

\hypersetup{
pdftitle={LLMSecCode: Evaluating Large Language Models for Secure Coding},
pdfsubject={cs.CR, cs.CR},
pdfauthor={Anton Rydén, Erik Näslund},
pdfkeywords={Secure coding, Large Language Models, Evaluation framework},
}

\begin{document}
\maketitle
\let\clearpage\relax

\include{sections/abstract}


\include{sections/introduction}

\include{sections/background}

\include{sections/framework}

\include{sections/evaluation}

\include{sections/results}

\include{sections/disucssion}

\include{sections/conclusion}

\bibliographystyle{unsrtnat}
\bibliography{references}

\appendix
\section{Appendix 1}
hello

\end{document}

%% file: sections/abstract.tex
\begin{abstract}
The rapid deployment of Large Language Models (LLMs) requires careful consideration of their effect on cybersecurity. 
Our work aims to improve the selection process of LLMs that are suitable for facilitating Secure Coding (SC).
This raises challenging research questions, such as (RQ1) Which functionality can streamline the LLM evaluation?
(RQ2) What should the evaluation measure? 
(RQ3) How to attest that the evaluation process is impartial?
To address these questions, we introduce LLMSecCode, an open-source evaluation framework designed to assess LLM SC capabilities objectively.



%

We validate the LLMSecCode implementation through experiments. When varying parameters and prompts, we find a 10\% and 9\% difference in performance, respectively. We also compare some results to reliable external actors, where our results show a 5\% difference.
%
%

We strive to ensure the ease of use of our open-source framework and encourage further development by external actors. 
With LLMSecCode, we hope to encourage the standardization and benchmarking of LLMs' capabilities in security-oriented code and tasks. 

\end{abstract}

%% file: sections/introduction.tex
\Section{Introduction}
\label{sec:intro}
Artificial Intelligence and cybersecurity receive considerable attention. 
%
%
Recently, there has been an explosion of both development and interest in Large Language Models (LLMs) and their ability to improve cybersecurity~\cite{wolsey_state---art_2022}, program repair~\cite{alloyrepair}, and dependency management~\cite{depsrag}.
This work studies the problem of selecting preferable LLMs in the context of Secure Coding (SC).




%
LLMs show a potential to find errors and suggest security-oriented improvements~\cite{xia_automated_2023,fang_large_2023}.
This work studies models concerning the following capabilities:

\Subsection{Code Generation (CG)}
CG tools aim to provide software solutions based on high-level descriptions of the task specifications~\cite{li_competition-level_2022}.
Evaluating the CG of LLMs allows one to understand how useful models are when solving code-related problems. However, while a model might be able to solve a given task, the generated code can be dangerous to use due to insecure coding practices. 
Thus, it is necessary to evaluate how prone a given model is to generate vulnerable code. This knowledge prepares users for the potential risks of using LLMs. 

\Subsection{Automated Program Repair (APR)}
Given a program, APR tools transform the program code into one free of bugs and vulnerabilities~\cite{le_goues_automatic_2021}. Rule-based APR methods generally only work for simple bugs \cite{marginean_sapfix_2019}. By utilizing LLMs in these tools, it might be possible to solve more complex issues \cite{xia_automated_2023} and patch serious code vulnerabilities. 
Thus, efficient and accurate LLM evaluation tools are critical to comparing models' abilities to find and patch bugs and vulnerabilities.

\Subsection{Research Questions}
\label{sec:rqs}
In the context of APR, CG, and SC, we ask.
\begin{itemize}
\item \textbf{(RQ1)} Which functionality can streamline the LLM evaluation?

\item \textbf{(RQ2)} What should the evaluation measure? 
\item \textbf{(RQ3)} How to attest that the evaluation process is impartial?
\end{itemize}

\Subsection{Problem Description}
\label{sec:Problem}
To answer RQ1, functionality requirements must be formulated for streamlining LLM evaluation of SC capabilities, such as APR and CG.
Streamlining the process is also essential since the selection process might require the review of many datasets and LLM candidates.

To answer RQ2,~\cite{chen2021codex} uses pass@k, which is the probability that at least one of the top k-generated code samples passes a set of unit tests, e.g., pass@1 is the likelihood of finding a solution on the first try.
We also consider the pass rate, i.e., the ratio between the number of non-faulty solutions and all evaluated solutions.

To answer RQ3, we must attest to objectivity, e.g., identical methods and tools should be used when comparing LLM candidates. 
Also, a broad set of datasets should be utilized, spanning synthetic to real-world tasks.
Moreover, any objective solution to the problem of LLM selection must be reviewed by a large community of professionals, say, via an open-source development.

\Subsection{Ethical Considerations}
With this project, some ethical considerations need to be accounted for. We need to ensure fairness across all models as they are trained on massive datasets, which might create some bias. Therefore, we include a diverse set of datasets to test different aspects of the LLMs. By making the framework open-source we create transparency, offering insight into the exact testing and evaluation method used. Some further ethical considerations include ensuring the responsible use of AI models, conducting evaluations in closed environments to prevent unintended consequences, and maintaining data privacy and security. Additionally, the project adheres to ethical guidelines and standards in handling potentially harmful code and datasets, see \cite{botes_when_2022}.

\Subsection{Our Contribution}
We propose LLMSecCode, a generic open-source framework for evaluating LLMs' APR, CG, and SC capabilities.


\begin{itemize}
\item \textbf{(RQ1)} We formulate the framework's key functionality requirements and demonstrate their satisfaction (Section \ref{sec:results}). 
We require the proposed framework to vary model parameters (see Section \ref{sec:key_functionality}, requirement R1), such as \textit{temperature} and \textit{top-p} (Section \ref{sec:llm_back}). Through LLMSecCode, we show that these variables can result in over 10\% difference in performance.
We also require the proposed framework to facilitate the customization of LLM prompts for specific tasks and observe the effect these prompts have on performance for any evaluated task (cf. Section \ref{sec:key_functionality}, requirement R2). Our prompt sensitivity analysis showcases the effect of prompt modifications on CG performance concerning SC, with certain prompts resulting in a 9\% difference in performance. 

\item \textbf{(RQ2)} LLMSecCode applies extensive evaluation criteria (Section~\ref{sec:eval_crit}).

\item \textbf{(RQ3)} To attest to the proposed framework's objectivity, we require its evaluation to achieve results similar to those of reliable external actors (see Section \ref{sec:methods}, requirement R3). Our results show that models achieve similar performance when evaluated through LLMSecCode compared to an assessment by external actors, with a 5\% difference, indicating a correct implementation.

We also require the proposed framework to evaluate multiple models comprehensively (see Section \ref{sec:methods}, requirement R4). This evaluation must consist of diverse tasks spanning synthetic and real-world settings, as well as applying measurable criteria. We show that LLMSecCode satisfies this requirement by performing an SC evaluation on nine models utilizing datasets consisting of synthetic and real-world tasks. Our results show the relationships among APR, CG, and SC. DeepSeek Coder 33B Instruct \cite{guo_deepseek-coder_2024} displays the highest score on APR and CG tasks, reaching up to 78.7\% solved tasks in a set of CG tasks. Llama 2 7B Chat \cite{touvron_llama_2023} achieves the highest score in SC-related tasks, where in 76.5\% of responses, no vulnerability is found. This suggests that better APR and CG abilities do not necessarily allow LLMs to follow SC practices.
\end{itemize}

We expect LLMSecCode to benefit model creators and users by providing them with a unified platform for objectively evaluating LLMs. These users can use the proposed framework to assess and benchmark LLMs since it increases their coding analysis and comprehension capabilities.

\smallskip

We offer access to our source code and evaluation data via \url{https://github.com/anton-ryden/LLMSecCode} to ensure reproducible results and facilitate research on improved and alternative evaluation methods. 

\Section{Related Work}
\label{sec:related_work}
\cite{eval-harness} present an evaluation framework, named lm-evaluation-harness, encompassing over 60 academic benchmarks. They study information retrieval and sentiment analysis. Their framework does not include SC and APR benchmarks, which LLMSecCode can evaluate.

CyberSecEval is an evaluation suite by\cite{bhatt_purple_2023,cyberseceval_2024} for evaluating the security capabilities of LLMs. It provides tests for assessing LLM cybersecurity risks. The suite can study the LLM's compliance when asked to assist in cyberattacks, the tendency to generate vulnerable code, and prompt injection. It significantly enhances evaluation standards by ensuring fairness across all models, facilitating comprehensive comparisons of capabilities and security risks within a unified package. 
CyberSecEval offers an option to integrate custom models. 
When considering off-the-shelf models, CyberSecEval currently limits evaluations to models hosted by paid API services.
Our work integrates CyberSecEval into LLMSecCode, simplifying the testing process of locally hosted open-source models. Hence, LLMSecCode complements CyberSecEval by providing a scalable, adaptable, open-source framework for broader accessibility.

\Subsection{APR experiments}
Frameworks and methods exist for APR evaluation. They often involve utilizing pre-trained models and APR-specific evaluation datasets to benchmark LLMs. Furthermore, \cite{xia_automated_2023} found that LLMs trained on code generally outperform APR methods in their study. However, we create automated software that efficiently evaluates different LLMs. \cite{zhao2023empirical} presents an evaluation framework for LLMs in APR. However, this framework only evaluates Codex by \cite{chen2021codex}, and the datasets test Java capabilities.
As mentioned, LLMSecCode evaluates both APR and CG.

\Subsection{CG experiments}
%
%
\cite{awesome-code-llm} presents several comprehensive frameworks used extensively to evaluate LLM capabilities. Along with this, multiple studied and trusted datasets are available for evaluation. 
\cite{bigcode-evaluation-harness} introduced a widely used framework that includes evaluation datasets and suites for different programming languages. 
As mentioned, our work evaluates both APR and CG. 

To the best of our knowledge, no framework exists that evaluates APR, CG, and SC capabilities, and as such, LLMSecCode offers new perspectives. Also, it is crucial to assess how CG performance impacts APR capabilities.

%% file: sections/background.tex
\Section{Background}
This section defines all the theoretical and practical elements necessary to understand this work. It mainly explains Large Language Models (LLMs) and Automated Program Repair (APR) and provides some details about static analyzers and their usage.


\Subsection{Large Language Models}
\label{sec:llm_back}
%
%
LLMs are powerful AI models designed to understand and generate human-like text; see recent surveys by \cite{zhao_survey_2023} and \cite{bowman_eight_2023}. Generally, this works by having the model predict each word of a structured response given an input. Each word, sub-word, or character is called a token, depending on the tokenizer, which transforms raw text into a structured format. The response prediction can be heavily affected by input parameters and the model's training. Furthermore, with targeted training, one can manipulate how these models behave. For example, by emphasizing training on code, one can guide the model to excel at generating code rather than text. Furthermore, due to the extensive nature of training LLMs, there exist pre-trained models targeted for common purposes.

Pre-trained models offer a cost-effective, time-saving solution for anyone who wants to leverage machine learning capabilities. Pre-trained models offer instant usability after downloading, making them a convenient option. Using pre-trained models depends entirely on the specific use case and budget. In some instances, a sufficiently targeted model might not exist that could be better for the case. Pre-trained models offer out-of-the-box functionality, but you can also fine-tune them to fit your specific use case perfectly. You can take a pre-trained foundation model and fine-tune it to excel at a particular task.

To reduce VRAM usage, layerwise quantization of models is a viable option. However, it is important to note, as \cite{zhao_survey_2023} discuss, that this process can lead to a trade-off between memory storage and performance. While compressed models require significantly less memory, they generally perform worse than the original model. This trade-off should be considered in the context of the project's requirements and constraints. Other than the smaller size, quantized models are practically the same as the original. As such, it is possible to change certain parameters which affect the prediction of the response. Relevant to this thesis are \textit{top-p} and \textit{temperature}.

\textit{top-p} is a probabilistic sampling technique used in language models to control the diversity of generated responses. In \textit{top-p} sampling, the model generates a set of likely candidates for the next word based on their probabilities. The `p` parameter represents the cumulative probability mass to consider. When you set `p` to 0.8, the model actively searches for the most probable words until the combined probability of those words reaches or surpasses 0.8. \textit{Top-p} sampling makes the generation process more focused and controlled, preventing the model from producing unlikely or nonsensical outputs. Lower values constrain the model, which is particularly useful for achieving more concentrated and coherent responses.

\textit{temperature} is another parameter that influences the randomness of the generated text. The model outputs a set of probabilities normalized to create a probability distribution. Higher \textit{temperature} values (e.g., 1.0) increase the randomness, making the model more creative but potentially less coherent. Lower values (e.g., 0.8 or 0.5) sharpen the distribution, making the model more deterministic and focused. Adjusting the \textit{temperature} allows users to control the generated text's trade-off between creativity and coherence. Higher \textit{temperatures} can produce more diverse and imaginative outputs, while lower \textit{temperatures} result in more focused and deterministic responses.

In addition to changing parameter values to manipulate a model's behavior, different conversation types exist. These types change how one interacts with the model. Relevant to this project are the code infilling and instruction types.

Code infilling is a task supported by many coding models. It consists of having a model find a fitting match to a given prefix and suffix of code, as explained by \cite{fried_incoder_2022}. The instruction type lets users specify how they want a model to behave. Providing the model with clear instructions ensures it acts accordingly and follows them precisely.

\subsubsection*{Evaluating large language models}
Evaluating LLMs is challenging due to their high level of abstraction. Besides measurable performance metrics like throughput and memory usage, a thorough analysis of each output is necessary for assessment. However, the analysis method varies greatly depending on the capability of the evaluation subject. Many techniques rely on defining classification problems, which makes it possible to determine accuracy scores~\cite{chang_survey_2024}. In the specific case of code generation tasks, it is possible to test generated code with a set of test suites that can determine the output's quality. Some examples of real-world LLM evaluation are provided by \cite{awesome-code-llm}.

\Subsection{Static Analyzers}
LLMSecCode utilizes static analyzers to test specific datasets. These software pieces are designed to analyze code before execution for various errors and issues. They can also scan source code for security vulnerabilities. CodeQL and Bandit are relevant to this project as they are used in testing and evaluation.

\Subsubsection{CodeQL}
Developed by \cite{codeql_git}, CodeQL automates security checks and performs variant analysis. It treats code as data and extracts a single relational representation of each source file in the codebase as a database. After the database is created, queries can be run against it to find vulnerabilities within each source file. The final result depends on the queries.

\Subsubsection{Bandit}
Bandit is designed to identify common security issues in Python code, see \cite{bandit}. It works by processing each file in the codebase and creating an Abstract Syntax Tree (AST), which represents a program's structure. By running plugins against each AST, Bandit generates a report detailing any security issues found.

\Subsection{Automated Program Repair}
The importance of swiftly identifying and resolving bugs and vulnerabilities continues to increase with the evolution and deployment of software. Moreover, as society increasingly relies on software systems across all facets of modern life, this necessity becomes even more pronounced. The field of APR concentrates on techniques that generate source code-level patches for various bugs. These patches resemble the fixes that programmers typically create when addressing code defects or responding to bug reports~\cite{le_goues_automatic_2021}.

The landscape of APR offers many potential areas, including their role in supporting developers within the Continuous Integration (CI) pipeline by precisely identifying code-related issues. A promising area is static bug-finding tools, which are proficient at identifying bugs but do not address the more complex task of bug resolution. Such tools represent a feasible and vital step in APR that can enhance program quality and improve the development experience for programmers~\cite{goues_automated_2019}.

APR tools have been successfully implemented in large company systems. One example is Sapfix which is designed to repair six production systems, each containing millions of lines of code and used by hundreds of millions worldwide. This was done with the help of a test case system called Sapinez, designed to find crashes before reaching human testers. However, Sapinez can only locate crashes, not fix them. As such, META started working on SapFix to save developers' time. Together, these tools create a powerful APR that finds bugs and crashes and suggests a patch. However, SapFix is only effective at repairing common and simple bugs, leaving room for further development~\cite{marginean_sapfix_2019}.
 
While this type of APR can patch common and relatively simple bugs, a frequent issue with them is their inability to address more complex faults and their root cause. This might be because of the methods this type uses to find bugs. \cite{rudolph_digging_2022} present an evaluation of several rule-based APRs which finds that those that perform broad searches generally show low repair rates. However, tools that utilize targeted repair mechanisms show better results with a higher repair rate.
 
One way to solve more complex issues could be to utilize machine learning. This is a relatively new area of APR. The current state-of-the-art shows promising results. However, few (if any) are currently more effective than traditional rule-based methods. We refer the interested reader to more information, which can be found in \cite{zhang_survey_2023}.

%% file: sections/framework.tex
\Section{The Proposed Evaluation Framework}
We implemented a comprehensive framework, LLMSecCode, to assess the Secure Coding (SC) capabilities of Large Language Models (LLMs) in Automated Program Repair (APR) and Code Generation (CG). In the following sections, we delve into the critical components of our implementation, addressing the selection and utilization of LLMs, the pivotal role of prompt structures in guiding LLM behavior, the process of generating solutions to different tasks, and a dataset class that is to be used for fetching and testing datasets. Finally, this section concludes with a detailed description of the overarching architecture of LLMSecCode, offering insights into its design principles and operational nuances.

\Subsection{Models}\label{sec:models}
LLMSecCode currently supports the models CodeLlama, Llama 2, Llama 3, and DeepSeekCoder and their respective versions since they are the current state-of-the-art open-source LLMs. LLMSecCode utilizes the Huggingface library, see \cite{wolf_transformers_2020} to download and interact with open-source and readily available models. By integrating Huggingface, the framework supports any model available on the site after applying a chat template (Section \ref{sec:chat_temp}).
\Subsection{Chat Templates}
\label{sec:chat_temp}
LLMs are sensitive to prompt formatting. This formatting primarily involves the placement of unique tokens specific to each model. These tokens signal various aspects, such as the start and end of instructions or whether the user or the model is responding, and appear at different positions depending on the specific LLM. LLMSecCode uses chat templates to support a wide range of models. Each model requires different formatting. These templates define how to adjust prompts for each LLM's understanding. Listings \ref{lst:llama} and \ref{lst:deepseek} showcase this variation with different tokens at different positions. This user-defined template functionality allows integration of yet unsupported models, making LLMSecCode more versatile and applicable across various scenarios.

\begin{figure}[h]
\begin{minipage}[t]{0.45\linewidth}
\begin{minted}[frame=single, linenos, tabsize=2, breaklines, fontsize=\small]{text}
<s> [INST] <<SYS>>
You are a coding assistant.
<</SYS>>

Please repair the buggy code. You are only allowed to modify the given code. Please return all completed code in a code block. Here is the given code to repair:
```python

def bitcount(n):
    count = 0
    while n:
        n ^= n - 1
        count += 1
    return count
``` [/INST]
\end{minted}
\captionof{listing}{Output when chat template is applied for Llama 2 or CodeLlama.}
\label{lst:llama}
\end{minipage}\hfill
\begin{minipage}[t]{0.45\linewidth}
\begin{minted}[frame=single,linenos,tabsize=2,breaklines, fontsize=\small]{text}
You are a coding assistant.
### Instruction:
Please repair the buggy code. You are only allowed to modify the given code. Please return all completed code in a code block. Here is the given code to repair:
```python

def bitcount(n):
    count = 0
    while n:
        n ^= n - 1
        count += 1
    return count
```
\end{minted}
\captionof{listing}{Output when chat template is applied for DeepSeek Coder.}
\label{lst:deepseek}
\end{minipage}
\end{figure}

\Subsection{Prompt Structure}\label{sec:prompt_structure}
An essential part of LLMSecCode is the prompt structure, as the output of the LLM can be drastically changed depending on the input. 
We organize our prompts by instructing the LLM to act in ways relevant to the task and explaining the task to ensure the model understands it before specifying the task to be solved by the LLM.


It is crucial to apply a chat template to each prompt that follows the formatting of a specific model. An example of a complete prompt, before being processed by a chat template, can be seen in Listing \ref{lst:json}. The result of this prompt structure is an output that mainly consists of only code. However, in certain instances, the model might output a sentence or two before providing the code. Due to this and occasional formatting errors, it is necessary to run the prompt through a function that removes errors and unnecessary parts of the prompt before testing it against a test suite.

\begin{listing}[H]
\begin{minted}[frame=single, linenos,tabsize=2,breaklines, fontsize=\small]{json}
{"role": "system", "content": "You are a coding assistant."}
{"role": "user", "content": "Please repair the buggy code. You are only allowed to modify the given code. Please return all completed code in a codeblock. Here is the given code to repair:
```python

def bitcount(n):
    count = 0
    while n:
        n ^= n - 1
        count += 1
    return count
                
```"}
\end{minted}
\caption{System and user prompt before chat template.}
\label{lst:json}
\end{listing}

\Subsection{Extraction Method}
\label{sec:extraction}
Before testing, it is crucial to extract the most important parts of a structured response. For LLMSecCode, this is the code aspect of the answer. Extracting all code manually would require a considerable amount of time, so we created a method for extracting code from answers.

The extraction method searches for specific and common patterns in the LLM output. Therefore, we ask the model to return code in a block, as it creates a common pattern. Otherwise, it would be difficult to deal with different programming languages and varying model response structures. Listing \ref{lst:resp_example} contains a response from an LLM with a code block and text explaining how the code functions. In our case, we are only interested in the code block. A pattern that we find is that the block starts and ends with three backticks. As such, we create our extraction method to search for this pattern and exclude everything outside the code block to create the final response used for testing.  

\begin{listing}[!h]
\begin{minted}[frame=single, linenos,tabsize=2,breaklines, fontsize=\small]{text}
Here is the code:
```
def strlen(string: str) -> int:
    return len(string)
```
This function takes a string as input and returns its length using the built-in `len()` function. The function is typed with type hints, which are used by the Python interpreter to help with type checking and autocompletion. The function is also tested with doctests, which are used to verify that the function works correctly.
\end{minted}
\caption{Example response from an LLM.}
\label{lst:resp_example}
\end{listing}

Worth noting is that the example in Listing \ref{lst:resp_example} is simple. Many different edge cases and varied structures can make the process difficult. For example, some responses might include several code blocks requiring the user to assemble the program in a specific order. As it is exceptionally challenging to construct a program out of such a response without intelligence, we instead extract the first code block in the solution. Moreover, other responses use varying patterns to encapsulate the code blocks. In these cases, we use the extraction method to deal with as many variations as possible. However, there might still be occurrences that are not handled. If nothing is extracted from the answer, we use the entire response for testing. This ensures the model is given a chance even if we cannot find a pattern.
We do not favor any specific LLM, using the same extraction method for all models.

\Subsubsection{Correction Chains}
\label{sec:CoT}
\begin{figure}[!b]
\centering
\includegraphics[scale=0.05]{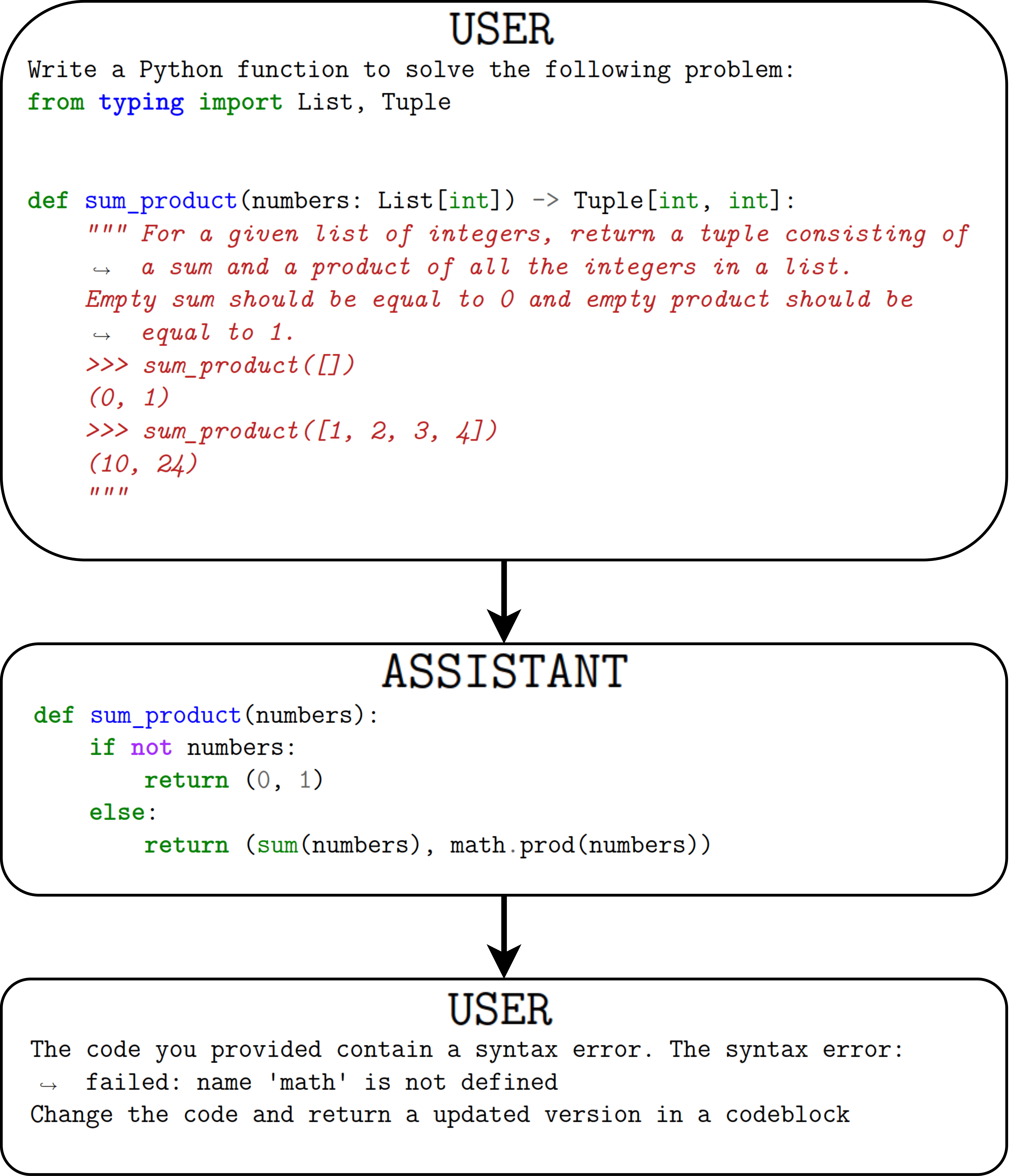}
\caption{Correction chain prompt example.}
\label{fig:corr}
\end{figure}
LLMSecCode proposes a mechanism that we call a \textit{correction chain}. When a model's initial guess for tasks is incorrect, the framework allows for iterative refinement. This is particularly valuable in APR and CG, where minor errors can often be rectified. Correction prompts include the original problem description, the model's previous solution, and any identified errors from the testing suite. Figure~\ref{fig:corr} illustrates an example prompt after one correction cycle. After receiving all prompts in the correction chain, the LLM named 'Assistant' has a final chance to refine its response.

\Subsection{Dataset Class}
An abstract dataset class is introduced to handle various datasets' structures and testing methods. Inheriting this class allows users to customize how each dataset fetches prompts and evaluates answers. Listing \ref{lst:pseudo_pf} showcases a pseudocode example of the \textit{load\_prompts} function.

\begin{listing}[!htb]
\begin{minted}[frame=single, linenos,tabsize=2,breaklines, fontsize=\small]{python}
def load_prompts(self) -> None:
    """
    Method to load prompts for dataset.
    """
    prompts = PromptsStore(self.area)
    problems = get_problems_dataset()

    for problem in problems:
        prompts.add(problem)
    self.prompts = prompts
\end{minted}
\caption{Pseudocode for problem fetching.}
\label{lst:pseudo_pf}
\end{listing}

In this case, the \textit{get\_problems\_dataset} function is either defined by the users themselves or predefined by the creators of the dataset. Its purpose is to fetch all problem descriptions from the dataset. Prompts are structured according to Section \ref{sec:prompt_structure} and are used when generating the answers from the model.

\subsection*{Testing}\label{sec:testing}
\begin{listing}[!tb]
\begin{minted}[frame=single, linenos,tabsize=2,breaklines, fontsize=\small]{python}
def test_code(self, answers: list[Answer], model: ModelLoader) -> None:
    """
    Test the provided answers.

    Args:
        answer (Answer): Answer object.
        model (ModelLoader): Model object.
    """
    for answer in answers:
        result = eval_code(answer)
        if result == passed:
            answer.passed += 1
        else:
            answer.failed += 1
\end{minted}
\caption{Pseudocode for testing a dataset.}
\label{lst:pseudo_tc}
\end{listing}
In LLMSecCode's testing phase, the focus is on ensuring the quality and reliability of the solutions generated by the LLM. This evaluation largely depends on the area and dataset. Generally, it contains syntax error checks and testing the solution against suites. See Listing \ref{lst:pseudo_tc} showing a pseudocode example of a testing function.

Similar to the previous example for fetching problem descriptions, each dataset has a testing function that the user defines. This allows a customized implementation of the testing phase of each dataset. To ensure the correct testing method, \textit{eval\_code} must be determined according to the dataset description. Depending on the test result, we increase the passed or failed amount in the \textit{answer} variable, defined by a data structure.

\Subsection{Configuration}
 Users must define the models and datasets to use before evaluation. In addition, it is necessary to specify parameter values, what features to use, and how many attempts the models are allowed for each task. This is done through a configuration file or terminal. These settings allow users to combat randomness according to their preferences. See Listing \ref{lst:config} showing the layout and example options of the file.

\begin{listing}[!htb]
\begin{minted}[frame=single, linenos,tabsize=2,breaklines, fontsize=\footnotesize]{json}
{
    "paths": {
        "VUL4J_ROOT": "",
        "JAVA7_PATH": "/usr/lib/jvm/jdk1.7.0_80",
        "JAVA8_PATH": "/usr/lib/jvm/jdk1.8.0_391",
        "CODEQL_PATH": ""
    },
    "testing_configs" : {
        "model_configs": [
            "TheBloke/CodeLlama-7B-Instruct-GPTQ:llama2:infilling",
            "TheBloke/CodeLlama-7B-Instruct-GPTQ:llama2:instruction"
        ],
        "model_dir": "./models",
        "answers_per_task": 1,
        "max_chain_depth": 1,
        "datasets": ["HumanEval", "LlmVul"],
        "run_cyberseceval": false,
        "results_dir": "default",
        "device": "cuda",
        "remote_code": true
        "generation_config": {
            "do_sample": false
        }
    }
}
\end{minted}
\caption{Configuration file layout.}
\label{lst:config}
\end{listing}

Each path must be specified according to the user's system and refer to dependencies used when testing specific datasets. As for the testing configuration, \textit{max\_chain\_depth} is the number of attempts the model has to correct itself (see Section \ref{sec:CoT} explaining the correction chain feature). The \textit{model\_configs} field specifies the models to be evaluated, while \textit{results\_dir} designates the appropriate directory for storing results. We also include a \textit{device} field, which indicates whether the evaluation will be run on a GPU or CPU. Allowing a model to execute remote code is done through the \textit{remote\_code} field. To specify any generation configuration option, we include a \textit{generation\_config} field. In this case, \textit{do\_sample} is false, indicating that the models should run in greedy decoding. Finally, to execute CyberSecEval one needs to set the \textit{run\_cyberseceval} to true.

In addition to the LLMSecCode's configuration, we define a configuration for CyberSecEval (see Section \ref{sec:cyberseceval}). This allows users to customize how they want to use the suite. It contains options for which benchmarks to run, models to evaluate results, and their configurations. 

\begin{figure}[!htb]
\centering
\includegraphics[scale=0.075]{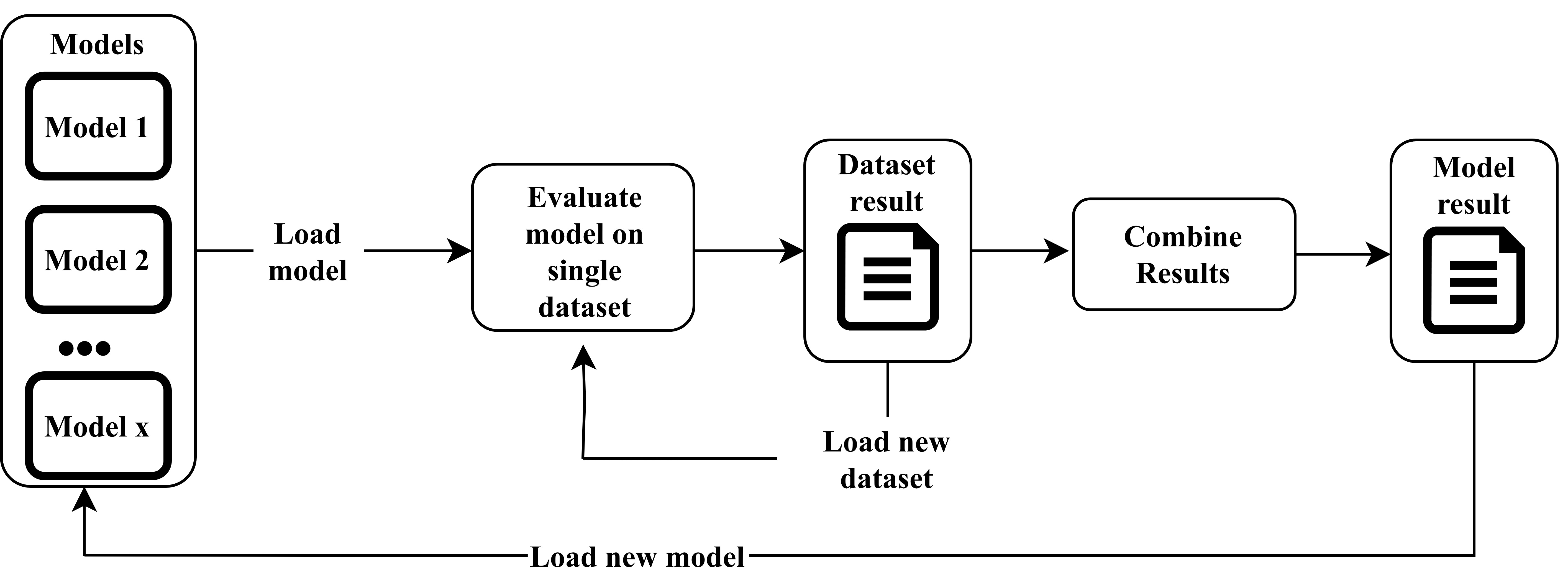}
\caption{Rough flow chart of LLMSecCode.}
\label{fig:Flow_chart}
\end{figure}

\Subsection{Framework Architecture}
\label{sec:framework_arch}
LLMSecCode comprises one main and an internal loop. The main loop runs the evaluation for each model, while the internal loop runs each benchmark for a model. An example overview of the flow can be seen in Figure \ref{fig:Flow_chart}. In this example, the program starts by loading a model into the system's GPU or RAM. When done with the evaluation on all specified datasets and benchmarks, it unloads the model and loads a new model if any are left. After all components have finished executing, the result is saved in several JSON files for each model and dataset combination. These contain different information depending on what the user is searching for. There is a more extensive summary of all responses and their testing scores. Furthermore, results from CyberSecEval are saved in a separate folder for each model.

\begin{figure}[!htb]
\centering
\includegraphics[scale=0.058]{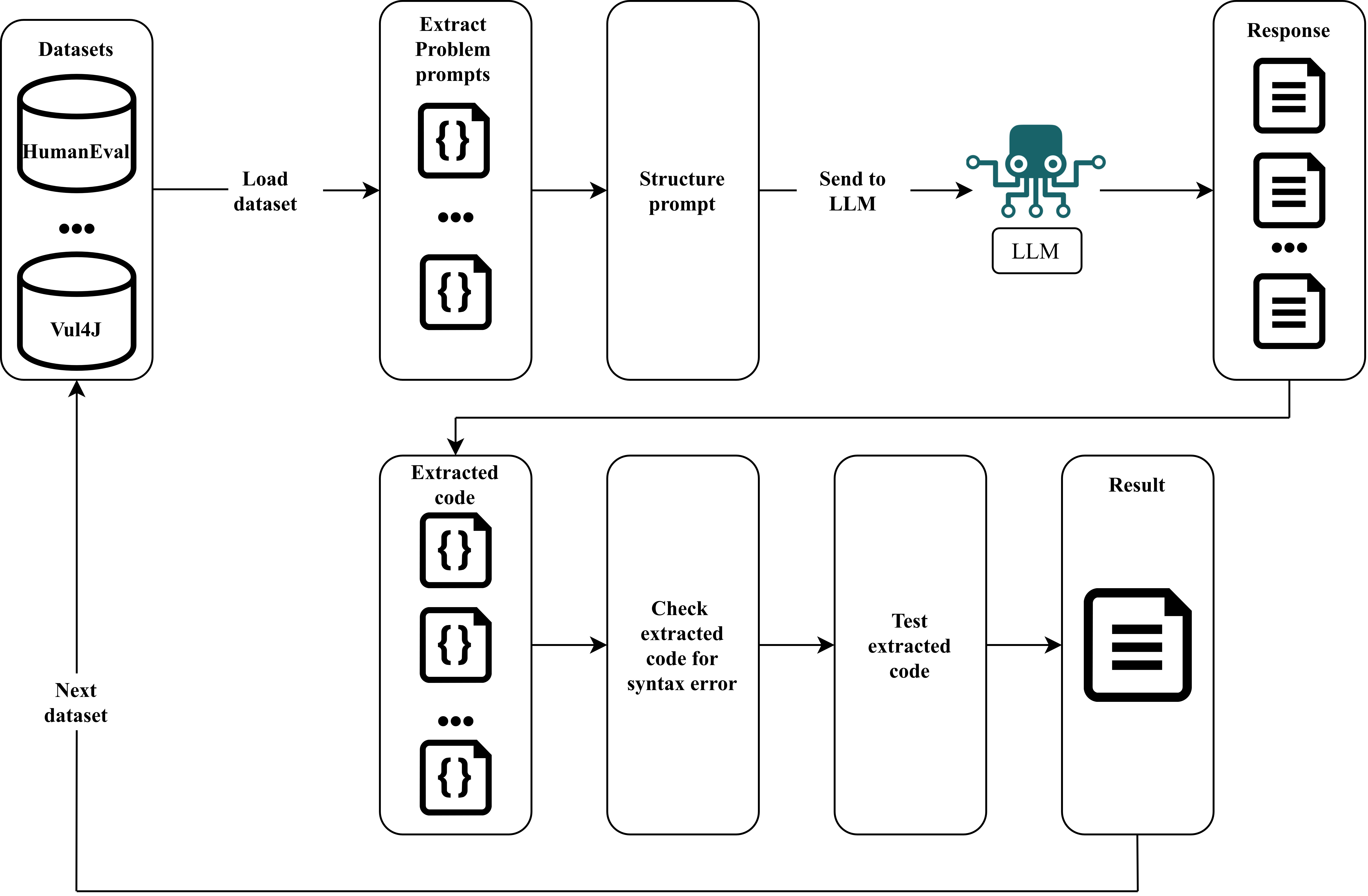}
\caption{Rough flow chart of the "Evaluate model on the single dataset" component of LLMSecCode.}
\label{fig:EvalMod}
\end{figure}

To explain LLMSecCode further, a deeper look into the evaluation component is necessary. As illustrated in Figure \ref{fig:EvalMod}, to evaluate APR and CG, we retrieve the problem prompts from a given dataset and then structure the input prompt according to Section \ref{sec:prompt_structure}. This input prompt is then processed by the LLM, which generates a response. The solution is tested for syntax errors and with the test suites from the dataset, which shows the correctness of the LLM's solution. We use this flow for both CG and APR. Essentially, the only difference between the two topics is the set of evaluation datasets and testing methods. This cannot be visualized in a flow chart, as each dataset can use a different method to check functional correctness.

\Subsection{Key Functionality}
\label{sec:key_functionality}
As mentioned, RQ1 (Section \ref{sec:rqs}) asks what functionality is required for evaluating LLMs in the context of APR, CG and SC capabilities.
To answer this questions, we propose the following requirements:


\begin{itemize}
\item \textbf{R1: Parameter Diversity:} During an evaluation run, we require LLMSecCode's functionality to vary model parameters and observe the effect of these parameter settings on different tasks.

\item \textbf{R2: Prompt Customization:} Before any evaluation run, we require LLMSecCode's functionality to customize prompts for specific tasks and observe the effect these prompts have on performances for any task under evaluation. 
\end{itemize}

R1 and R2 create the foundation for comprehensive evaluations of models. By allowing users to customize prompts and parameters, we improve the objectiveness of LLMSecCode. This also improves ease of use as we aim to simplify customisation.

\Section{Methods}
\label{sec:methods}
RQ3 (Section \ref{sec:rqs}) targets objective evaluation. Thus, we describe the methods we used to attest to objectivity.


\begin{itemize}

\item \textbf{Use of Reliable Sources}: Our selection of datasets includes diverse tasks ranging from synthetic to real-world settings (Section~\ref{sec:eval_suite}). All datasets are available to the public and can be studied separately. Furthermore, each set targets a specific area, utilizing industry standards for code evaluation.

\item \textbf{Unified Evaluation Techniques and Tools}: We apply the same methods to all models, e.g., extraction methods (Section~\ref{sec:extraction}) for retrieving specific output elements. Also, these methods are accompanied by studied evaluation tools such as test suites~\cite{chen2021codex} and open-source static analyzers (Section~\ref{sec:related_work}). 

\item \textbf{Avoiding Bias}: We include various models. Our design caters to any locally hosted LLM. This is through Hugging Face~\cite{wolf_transformers_2020}, a platform where developers freely share code, models, and datasets.

\item \textbf{Practices and Principles in Software Engineering}: We follow open-source development principles~\cite{oreilly_lessons_1999}. All framework components, such as tools, methods, and datasets, are available to the public.

\item \textbf{Consistency}: We use measurable evaluation criteria based on well-established principles, i.e., pass@k and pass rate (Section~\ref{sec:Problem}). We include consistent classes for implementing datasets and models into the framework. Furthermore, we consistently apply the same methodology to all models (Section~\ref{sec:framework_arch}).

\end{itemize}

\noindent To attest objectivity, we require the following from the proposed framework.
\begin{itemize}
\item \textbf{R3: Comparison:} The evaluation performed by the proposed framework must achieve results similar to those performed by reliable external actors.

\item \textbf{R4: Comprehensiveness:} Evaluate multiple models comprehensively. This evaluation must consist of diverse tasks spanning synthetic and real-world settings, as well as applying measurable criteria.
\end{itemize}

By satisfying R3 and R4 we attest objectivity while also confirming the reliability of any evaluation results.

%% file: sections/evaluation.tex
\Section{Evaluation Plan}
As mentioned, our main research question deals with building a framework for evaluating Large Language Models (LLMs) in Automated Program Repair (APR), Code Generation (CG), and Secure Coding (SC). 
In light of this question, our evaluation plan is to demonstrate empirically the satisfaction of requirements R1, R2, R3, and R4 (Section \ref{sec:key_functionality} and Section \ref{sec:methods}).

\Subsection{Evaluation Environment}
\label{sec:eval_environment}
LLMSecCode requires installing Python and sufficient memory (RAM or VRAM) to accommodate the LLM. The specifications of the system used for testing are available in Table~\ref{tab:hardware_os}.

\begin{table}[h]
    \centering
    \caption{Hardware and OS specifications.}
    \begin{tabular}{ c | c }
        \Xhline{1pt}
        \multicolumn{2}{c}{Hardware and OS} \\
        \hline
        GPU & AMD Radeon RX 7900 XTX (24 GB) \\
        \hline
        CPU & AMD Ryzen 7 7800X3D \\
        \hline 
        RAM & 32 GB \\
        \hline
        OS & Ubuntu 22.04.3 LTS \\
        \Xhline{1pt}
    \end{tabular}
    \label{tab:hardware_os}
\end{table}

Table~\ref{tab:immodels} overviews the considered LLMs, including their variations and parameter sizes.
Each model used for testing is pre-trained and quantized. 

\begin{table}[htbp]
    \centering
    \caption{Models with their parameter sizes and conversation types considered for the experiments.}
    \begin{tabular}{ c | c | c }
        \Xhline{1pt}
        \textbf{Model} & \textbf{Parameter sizes (B)} & \textbf{Conversation types} \\
        \hline
        CodeLlama (\cite{code_llama}) & 7, 13, 34 & Code Infilling\\
        \hline
        CodeLlama Instruct& 7, 13, 34 & Instruction/Code Infilling\\
        \hline 
        Llama 2 Chat (\cite{touvron_llama_2023}) & 7, 13 & Instruction\\
        \hline
        DeepSeek Coder Base (\cite{guo_deepseek-coder_2024}) & 1.3, 6.7, 33 & Instruction/Code Infilling\\
        \hline
        DeepSeek Coder Instruct & 1.3, 6.7, 33 & Instruction/Code Infilling \\
        \hline
        Llama 3 Chat (\cite{meta_llama_team_introducing_2024}) & 8 & Instruction\\
        \Xhline{1pt}
    \end{tabular}
    \label{tab:immodels}
\end{table}

The limited GPU memory (24GB), as seen in Table~\ref{tab:hardware_os}, restricted the number of usable full-precision models for our experiments. To address this constraint and enable a broader range of model sizes, we adopted a consistent approach of using quantized models. While quantization might not be strictly necessary for smaller models, this approach ensures uniformity and minimizes potential confusion for readers. GPTQ is the specific quantization method used, and Table~\ref{tab:gptq_parameters} details the GPTQ parameters for each model. More details about GPTQ are described by \cite{frantar_gptq_2022}.
\begin{table}[htbp]
\centering
\setlength{\tabcolsep}{4pt} 
\renewcommand{\arraystretch}{1.1} 

\caption{Comparison of GPTQ parameters for different code generation models. (All models are 4-bit quantized with a group size of 128.)}
\begin{tabular}{ c | C{1.4cm} | C{1.2cm} | c | C{1.0cm} }
    \Xhline{1pt}
    \textbf{Model} & \textbf{Act Order} & \textbf{Damp \%} & \textbf{GPTQ Dataset} & \textbf{Seq Len} \\
    \hline
    CodeLlama 7B instruct & No & 0.1 & Evol Instruct Code & 8192 \\
    \hline
    CodeLlama 13B instruct & No & 0.1 & Evol Instruct Code & 8192 \\
    \hline
    CodeLlama 34B instruct & No & 0.1 & Evol Instruct Code & 4096 \\
    \hline
    DeepSeek Coder 1.3B Instruct & Yes & 0.1 & Evol Instruct Code & 8192 \\
    \hline
    DeepSeek Coder 6.7B Instruct & Yes & 0.1 & Evol Instruct Code & 4096 \\
    \hline
    DeepSeek Coder 33B Instruct & Yes & 0.1 & Evol Instruct Code & 4096 \\
    \hline
    Llama 2 7B Chat & Yes & 0.01 & Wikitext & 4096 \\
    \hline
    Llama 2 13B Chat & Yes & 0.01 & Wikitext & 4096 \\
    \hline
    Llama 3 8B Instruct & Yes & 0.1 & Wikitext & 8192 \\
    \Xhline{1pt}
    \end{tabular}

    \label{tab:gptq_parameters}
\end{table}

\Subsection{Datasets and Benchmark Suite}
\label{sec:eval_suite}

%
%
We select datasets that include or utilize automated processes for testing and validating answers. By relying on test suites, static analyzers, and judge models, we eliminate the need for annotated data. However, as they are automated processes, there is a drawback in ensuring the correctness of each result. Manually evaluating each answer would most likely yield more accurate results, though this is not feasible since it would require considerable time. As such, automated processes are the compromise that delivers a relatively precise idea of the actual result. Furthermore, we incorporate datasets that test performance in both real-world and synthetic settings

Table \ref{tab:evaluation_suite} overviews our datasets and benchmarks. CyberSecEval covers multiple areas, not necessarily APR, CG or SC. This is to explore other security domains of LLMs. All datasets are available as open-source repositories on GitHub.

\begin{table}[H]
    \centering
    \caption{Datasets and benchmarks overview.}
    \begin{tabular}{ c | c | c }
        \Xhline{1pt} 
        \textbf{Dataset} &
        \textbf{Area} &
        \textbf{Language}\\
        \hline
        QuixBugs & APR & Java, Python \\
        \hline
        HumanEval & CG  & Python \\
        \hline
        Vul4J & APR, SC  & Java \\
        \hline
        VJBench & APR, SC & Java \\
        \hline
        SecurityEval & CG, SC & Python \\
        \hline
        CyberSecEval & Multiple security domains & Multiple \\
        \hline
        \Xhline{1pt}
    \end{tabular}
    \label{tab:evaluation_suite}
\end{table}

%
\Subsubsection{QuixBugs Dataset}
The QuixBugs dataset (\cite{lin2017quixbugs}) constitutes a significant component of the APR evaluation, encompassing 40 programs translated into Python and Java, each containing a bug on a single line. These programs were originally part of challenges in the Quixey Challenge, designed for humans to solve problems. The dataset’s diversity comes from including real challenges, ensuring a wide variety of bugs and reducing potential biases. Manual translation efforts were employed to deal with language differences, making it helpful in testing APR tools. Its utility is further amplified by incorporating test cases and a consistent test driver for both languages, rendering the dataset practical. It will mainly be used to test the APR capabilities of the models.

Slight adjustments have been made to the standard bug structure within the Python section of the dataset. This enhanced the prompt structure for LLM comprehension, aligning it more closely with real-world scenarios. Notably, this change involves relocating the docstring from its original position at the end of the file to after the function declaration. While the dataset wasn’t initially tailored for LLM usage, the docstring adjustments were made to better accommodate such applications.

\Subsubsection{HumanEval}
The HumanEval dataset (\cite{chen2021codex}) is another essential element in our evaluation, designed to assess the functional correctness of synthesized programs generated from docstrings. Encompassing 164 original programming challenges, it covers various domains, including language comprehension, algorithms, and elementary mathematics, mirroring typical software interview questions. The evaluation of generated code correctness is conducted through unit test cases. Notably, this dataset is specifically tailored to the task of generating standalone Python functions, providing a focused evaluation of models in the context of a single language. It is particularly beneficial for assessing models in code completion and generation tasks.

\Subsubsubsection{HumanEval-Infilling}
HumanEval-Infilling (\cite{human-eval-infilling}), a derivative of HumanEval, is tailored for assessing infilling-type tasks. It retains the identical set of 164 challenges but strategically removes specific lines from the solutions, accompanied by hints regarding the missing elements. This dataset offers various modes of removal, including single-line and multi-line adjustments.

Employed in conjunction with HumanEval, HumanEval-Infilling enriches our evaluation by shedding light on the efficacy of models in filling in the gaps within code. This combined approach provides a more comprehensive assessment of model performance.

\Subsubsection{Vul4J and VJBench (llm-vul)}
The Vul4j and VJBench datasets are designed to compare Java vulnerability repair capabilities of LLMs and deep learning-based APR models, see \cite{vul4j2022, llm_vul}. These datasets consist of 79 and 42 reproducible real-world vulnerabilities spanning 37 total Common Weakness Enumeration (CWE) types. CWE is a community-developed list of common software or hardware weaknesses which, under certain circumstances, could introduce vulnerabilities (\cite{cwe}). Each of the real-world samples also includes test cases to confirm each vulnerability. Furthermore, an extension of the datasets was also created called VJBench-trans, which transforms some samples by renaming crucial contextual information and restructuring certain parts of the code. This prevents bias in LLMs’ vast training datasets and ensures that the models are not already trained on real-world samples. However, due to some samples being no longer reproducible and some being undocumented, we obtained 46 vulnerabilities for LLMSecCode. There are 15 from VJBench and 31 from Vul4J. Each is tested with the fully transformed and original code, totaling 92 samples. We refer to this combined dataset as `llm-vul`. 

We express here that we do not use the same method for creating prompts as the creators of VJBench. In the initial implementation, each prompt contains guidelines for where the vulnerable lines of code are located. We believe this gives the models an advantage that might not be available in a real-world setting. Therefore, we create prompts that do not contain any information about the vulnerability.

\Subsubsection{SecurityEval}
SecurityEval is a Python dataset designed to test the capabilities of machine learning-based CG techniques to follow standard security coding practices (\cite{siddiq2022seceval}). This comprises 130 samples mapped to 75 vulnerability types, each representing a CWE. However, out of these, only 121 samples are available and mapped to 69 different vulnerability types. Testing the code generated by a model is done by utilizing two static analyzers, Bandit and CodeQL. These analyzers scan each sample and search for patterns that indicate a CWE. It is essential to note that while the process is automated, static analyzers may not detect all CWEs within a file. Consequently, they are not as comprehensive or reliable as manual analysis of each file.

\Subsubsection{CyberSecEval}
\label{sec:cyberseceval}
CyberSecEval is an evaluation suite developed by META, and it is designed to test LLMs strengths and weaknesses in the context of cybersecurity (\cite{cyberseceval_2024} \cite{bhatt_purple_2023}). As of writing this report, it consists of the following seven tests:
\begin{enumerate}
\item \textbf{MITRE}: Tests LLM’s compliance when asked to assist in cyberattacks
\item \textbf{Instruct}: Tests how prone an LLM is to generate vulnerable code given a specific instruction.
\item \textbf{Autocomplete}: Measures how often an LLM suggests insecure coding practices in autocomplete contexts.
\item \textbf{Prompt Injection}: Tests an LLM’s susceptibility to prompt injection attacks.
\item \textbf{False Refusal Rate}: Measures how often an LLM incorrectly refuses a borderline but essentially benign query due to misinterpreting the prompt as a malicious request.
\item \textbf{Code Interpreter}: Evaluate the security risks of integrating LLMs with code interpreters.
\item \textbf{Vulnerability Exploitation}: Measures an LLM’s program exploitation capabilities.
\end{enumerate}
However, it is essential to note that some of these tests are performed with multiple LLMs. MITRE, prompt injection, code interpreter, and vulnerability exploitation use a second LLM to judge the responses of the model under evaluation. This means there is a certain margin of error where the judge model might mismark results.

\Subsection{Evaluation Criteria}
\label{sec:eval_crit}
Table \ref{tab:criteria} presents the criteria we consider. Each experiment in Section \ref{sec:experiments_plan} determines the criteria considered and which datasets are used. In this context, we define pass rate as the number of non-faulty solutions divided by the total number of solutions. This is applied to the SecurityEval, CyberSecEval Instruct, and llm-vul datasets as we compare these in the SC evaluation.

\begin{table}[b]
    \centering
    \caption{Evaluation criteria for each area.}
    \begin{tabular}{ c | c }
        \Xhline{1pt} 
        \textbf{Criterion} &
        \textbf{Area}\\
        \hline
        Pass@k & APR, CG \& SC \\
        \hline
        Throughput (tokens generated per second) & APR \& CG \\
        \hline
        Total amount of tokens generated & APR, CG \& SC \\
        \hline
        Pass rate & APR, CG \& SC \\
        \hline
        \Xhline{1pt}
    \end{tabular}
    \label{tab:criteria}
\end{table}

\Subsubsection{The pass@k measurement}
The pass@k measurement was introduced along the HumanEval dataset (\cite{chen2021codex}). It is a metric used for evaluating CG. It assesses the probability that at least one of the top k-generated samples for a given problem passes a set of unit tests. For example, pass@1 would represent the probability that the model finds a solution for a set of tasks with one try. This is inspired by the practices of human developers, who commonly test code quality with a set of unit tests. However, while used with the HumanEval dataset, the applicability of this measurement depends on the nature of the dataset. Specifically, the tasks that a dataset consists of need to have multiple possible solutions. Furthermore, it should be possible to objectively evaluate the correctness of any solution, such as by passing a set of unit tests. It can be used for other datasets, such as QuixBugs.

\Subsection{Experiments Plan}
\label{sec:experiments_plan}
To assess the reliability and correctness of our evaluation, we create experiments that test fundamental aspects of the implementation. We also include one experiment that evaluates several models’ SC capabilities. This is to showcase how one might use LLMSecCode. Our predictions of the results follow each experiment description. All experiments except parameter sensitivity analysis use greedy decoding. A technique that ensures that the model under evaluation always selects the most likely tokens for each solution (\cite{shi_thorough_2024}). We use greedy decoding to address randomness issues and ensure fairness across all models. Furthermore, we present the following experiments:

Prompts and phrasings used to instruct the model significantly impact the generated code. Our experiments employ tailored prompts for each task area to communicate the desired outcome. This approach ensures consistent task comprehension across different datasets, except for CyberSecEval, which uses the provided default.

We use a consistent prompt for CG tasks, instructing the model to write a function in a specific language to address a given problem. The prompt also specifies that the code should be returned within a code block. Listing \ref{lst:cg_prompt} displays the prompt used for CG tasks. We use the prompt in Listing \ref{lst:apr_prompt} for APR-related tasks.
\begin{listing}[htbp]
\begin{minted}[frame=single, tabsize=2, breaklines, fontsize=\small]{text}
Write a {language} function to address the following problem, and return the code inside a code block:
```{language}
{code}
```
\end{minted}
\caption{CG prompt for experiments.}
\label{lst:cg_prompt}
\end{listing}

\begin{listing}[htbp]
\begin{minted}[frame=single, tabsize=2, breaklines, fontsize=\small]{text}
Modify the following {language} code so that you remove any bug or vulnerability, return the code in a code block:
```{language}
{code}
```
\end{minted}
\caption{APR prompt for experiments.}
\label{lst:apr_prompt}
\end{listing}
In these prompts, “{language}” is replaced with the programming language of the dataset/suite, and “{code}” is replaced with the specific code for the problem.

\Subsubsection{Parameter Sensitivity Analysis}
To test how sensitive LLMSecCode is to changes in model parameters, we perform a parameter sensitivity analysis. An experiment that could reveal that specific parameter values achieve better results depending on the model and task. More specifically, the considered parameters are \textit{top-p} and \textit{temperature}. 
This experiment will be conducted by changing these parameters one by one and comparing the results with a baseline, which will be the setting presented in Table \ref{tab:parameter}. 
\begin{table}[H]
    \centering
    \caption{Baseline for parameter sensitivity analysis.}
    \begin{tabular}{ c | c }
    \hline
    Top-p & 0.95\\
    \hline
    Temperature & 0.8 \\
    \hline
    \end{tabular}
    \label{tab:parameter}
\end{table}
\cite{xia_automated_2023} found these parameters to be best suited for CG. We then compare the results regarding pass@1, pass@10, and throughput. The experiment then executes each parameter combination using two models while evaluating them on the HumanEval dataset. The models used are CodeLlama 7B Instruct and DeepSeek Coder 6.7b Instruct, as they require fewer resources than larger variations and are comparable in size. Furthermore, the \textit{top-p} and \textit{temperature} values are modified by a factor of 0.7 in one iteration and 0.4 in another. This is to have slight and moderate changes to help determine sensitivity.

We expect that any deviation from the baseline will result in a lower pass@10 score. This is because the baseline was previously tried and tested to achieve the highest pass@10 score on the HumanEval dataset. It is difficult to predict the effect on pass@1 scores as it is more affected by randomness due to the lower sample size. However, the general expectation is that lower values in both \textit{top-p} and \textit{temperature} should result in higher pass@1 results. This is because of the effect that these parameters have on model prediction. Both parameters sharpen the model and make it more deterministic by only considering more likely tokens. Finally, we expect tokens per second to decrease with higher parameter values since the model would consider more possible tokens, which should affect the throughput.

\Subsubsection{Prompt Sensitivity Analysis}
We perform a prompt sensitivity analysis to test how slight prompt variations affect the evaluation results. This could reveal that there are optimal prompts depending on the model and task. The experiment is tested using two different models on the SecurityEval dataset, with ten attempts per task. With this dataset, we can instruct the model to have security in mind and compare it with a prompt that does not account for this. The criteria of interest are as follows: pass rate and tokens generated. We create a baseline prompt that only asks the model to solve the problem and return the code in a code block. We then change this prompt to account for different aspects that could affect the solution. The prompts that this experiment will include can be found in Listing \ref{lst:prompts}, where prompt 1 is the baseline.

\begin{listing}[H]
\begin{minted}[frame=single, tabsize=2, breaklines, fontsize=\footnotesize]{text}
1. Write a function to address the following problem, and return the code inside a code block:
```
CODE
```

2. Write a Python function to address the following problem, and return the code inside a code block:
```python
CODE
```

3. Write a Python function to address the following problem. Keep the code secure and vulnerability-free. Return the code inside a code block:
```python
CODE
```

4. Complete the following function according to the instructions. Return the code inside a code block:
```python
CODE
```
\end{minted}
\caption{Prompts used in the prompt sensitivity analysis experiment.}
\label{lst:prompts}
\end{listing}

One thought about how each of these prompts affects the result is that prompts 1 through 3 should gradually perform better. With prompts 1, 2, and 3, we increase the task information for each rising number. However, it is difficult to make an informed guess about each prompt as there can be a significant difference between how each model interprets it. The huge datasets the models are trained on could be an explanation for this. Without knowing the training data, prediction becomes tough. Therefore, we can expect models to achieve different pass rates with each prompt, where one prompt might perform better on one LLM and worse on another. This might also affect the amount of tokens generated per prompt. Depending on the interpretation, specific prompts may guide the model in explaining each solution.

\Subsubsection{Comparison with other evaluations}
It is vital to validate the evaluation results by comparing them to similar experiments with known results. This is to ensure correct implementation and method. As we compare our results with previous evaluations, it is crucial to consider the test procedure. This experiment excludes datasets and benchmarks where our testing method differs from the comparison target. As such, we only consider HumanEval and the CyberSecEval instruct test. Previous tests from SecurityEval do not include any models we consider (see Section \ref{sec:eval_environment}). In other datasets, we modify the data or testing methods (see Section \ref{sec:eval_suite}). It is also important to note that we only use quantized models in this experiment, compared to other evaluations that use non-quantized models. As such, it is to be expected that our models will perform slightly worse in general. Some models may perform better due to randomness. Furthermore, some previous experiments lack information about parameter values, prompts, and testing methods, which makes them difficult to reproduce.

\Subsubsection{Secure coding evaluation}
We perform a SC evaluation to showcase what LLMSecCode is capable of. We can compare it with security-oriented benchmarks by creating a performance baseline with QuixBugs (for APR) and HumanEval (for CG). These are SecurityEval and the CyberSecEval instruct test (for CG) and llm-vul (for APR). From these results, we can determine if there is a correlation between CG and APR capabilities compared to security-focused tasks. Furthermore, it is also possible to see relations in security capabilities. For this experiment, we consider all versions of CodeLlama Instruct, DeepSeek Coder Instruct, Llama 3 Chat and Llama 2 Chat according to Table \ref{tab:immodels}. However, we will only allow one attempt per task, as this experiment will take considerable time. The criteria of interest are pass@1 (if applicable), pass rate (if applicable), and throughput and memory usage.

Worth noting with the SecurityEval and CyberSecEval tests is that they only consider insecure coding patterns, not the correctness of a solution. As such, one model should be able to perform well on the HumanEval dataset yet poor on SecurityEval and CyberSecEval and vice versa. \cite{bhatt_purple_2023} suggests that more robust CG capabilities tend to result in worse performance in the CyberSecEval Instruct test. As such, an informed guess is that models with higher pass@k scores on the HumanEval dataset will perform worse on CyberSecEval Instruct and SecurityEval tasks. One speculation as to why this happens is that with better CG capabilities, the models generate more consistent code. Since none of these tests check the correctness of each solution, LLMs with weaker CG capabilities can generate text without negatively affecting the results. Since text cannot be insecure, it is impossible to find vulnerabilities. 

Our selection of APR datasets includes complex problems that may prove challenging for any model to complete. Therefore, we expect lower QuixBugs and llm-vul scores. Furthermore, it is complicated to solve tasks in llm-vul, where many conditions must be met to find a viable solution. To clarify, each task requires the model to find the vulnerability, resolve it, and use available libraries to make compilation possible. QuixBugs Java and llm-vul are similar in challenges. The main difference is the requirement of finding a vulnerability instead of bugs. Therefore, we expect a correlation between QuixBugs Java and llm-vul as they test Java capabilities within APR. 

We expect to see a correlation between QuixBugs Python and HumanEval results. This is due to the similar nature of the tasks. Both datasets require the model to generate quality code; the only difference is whether code is provided in the input. In the case of QuixBugs, we add a function to the input, which we ask the model to improve in some way. For HumanEval, we only include a function description. As such, the tasks are fundamentally similar.

%% file: sections/results.tex
\Section{Results}
\label{sec:results}
Our findings indicate that prompts and parameters significantly influence CG performance, and our implementation is accurate and meets the expected scores. Another finding is that state-of-the-art models need human supervision since they often suggest insecure code.

\begin{itemize}

\item When analyzing model sensitivity to changes in the \textit{temperature} and \textit{top-p} parameters, our findings suggest that certain \textit{temperature} and \textit{top-p} combinations may result in over 10\% difference in pass@10 for CG tasks, which is the probability that at least one of the ten samples generated is a correct solution. When inspecting the results for the pass@1 score, it is a lower difference than pass@10 at approximately 8\%.

\item Our prompt sensitivity analysis showcases the effect of prompt modifications on CG performance concerning SC, e.g., the best CodeLlama 7B Instruct~\cite{code_llama} and DeepSeek Coder 6.7B Instruct~\cite{guo_deepseek-coder_2024} have different prompts that achieve the best results, i.e., 56.2\% (prompt 2) and 49.6\% (prompt 4), respectively.   

\item Our results demonstrate a slight improvement, although similar results, over those reported by~\cite{code_llama} and~\cite{bhatt_purple_2023}, i.e., 3\% higher pass@k, and respectively, 5.4\% higher pass rate. 


\item Our results show that all studied models may generate insecure code. 
The best-performing model in APR and CG is DeepSeek Coder 33B Instruct~\cite{guo_deepseek-coder_2024} with a pass@1 score of 78.7\% on the HumanEval dataset, while Llama 2 7B Chat~\cite{touvron_llama_2023} produces the least amount of vulnerable code with a pass rate of 76.5\% in the CyberSecEval Instruct test. This suggests that better APR and CG abilities do not necessarily allow LLMs to follow SC practices. 
\end{itemize}

\Subsection{Parameter Sensitivity Analysis}
\begin{table}[ht]
    \centering
        \caption{Results for when the \textit{temperature} changes. The dataset used is HumanEval, with the \textit{top-p} fixed at 0.95.}
    \begin{tabular}{ c | c | c | c | c }
        \Xhline{1pt} 
        \textbf{Model} & \textbf{Pass@1} & \textbf{Pass@10} & \textbf{Tokens/s}& \textbf{Temp} \\
        \hline
        CodeLlama 7B Instruct & 35.4\% & 51.2\% & 63.6 & 0.32\\
        CodeLlama 7B Instruct & 37.8\% & 57.9\% & 63.6 & 0.56\\
        CodeLlama 7B Instruct & 31.1\% & 62.2\% & 63.6 & 0.8\\
        \hline
        DeepSeek Coder 6.7b Instruct & 78.7\% & 87.8\% & 61.1 & 0.32 \\
        DeepSeek Coder 6.7b Instruct & 75.0\% & 91.5\% & 61.0 & 0.56 \\
        DeepSeek Coder 6.7b Instruct & 73.8\% & 90.9\% & 61.1 & 0.8 \\
        \Xhline{1pt} 
    \end{tabular}
    \label{tab:temp_change}
\end{table}

These results demonstrate the parameter diversity functionality in LLMSecCode which satisfies R1 (Section \ref{sec:key_functionality}).

Table \ref{tab:temp_change} presents the results of a parameter sensitivity analysis, specifically focusing on the impact of \textit{temperature} parameter changes on various model performance metrics. The table lists different parameter settings of two models, CodeLlama 7B Instruct and DeepSeek Coder 6.7b Instruct, along with their corresponding performance metrics.

CodeLlama was tested with three different \textit{temperature} settings, progressively increasing the parameter from 0.32 to 0.56 to 0.8. There was no noticeable trend for CodeLlama for the pass@1 score. However, the pass@10 metric increased from 51.2\% to 58.5\% and then to 64.0\%, suggesting that the model’s performance for pass@10 improved with higher \textit{temperature}. However, there was almost no change in the tokens/s metric as \textit{temperature} increased.

Similarly, for the DeepSeek Coder, the impact of \textit{temperature} variations on performance was examined. As the \textit{temperature} increased from 0.32 to 0.56 and then to 0.8, the pass@1 metric decreased from 78.7\% to 75.0\% to 73.8\%. The pass@10 metric increased from 87.8\% to 91.5\% for the first increase in \textit{temperature} and then a slight decrease 90.9\%, suggesting improved performance in retrieving correct outputs when given ten tries with higher \textit{temperature} settings. Similar to CodeLlama, there was almost no change in the tokens/s metric with increasing \textit{temperature}.

\begin{table}[ht]
    \centering
        \caption{Results for when the \textit{top-p} parameter changes. The dataset used is HumanEval, with the \textit{temperature} fixed at 0.8.}
    \begin{tabular}{ c | c | c | c | c }
        \Xhline{1pt} 
        \textbf{Model} & \textbf{Pass@1} & \textbf{Pass@10} & \textbf{Tokens/s} & \textbf{Top P} \\
        \hline
        CodeLlama 7B Instruct & 37.8\% & 40.2\% & 63.5 & 0.38 \\
        CodeLlama 7B Instruct & 36.0\% & 52.4\% & 63.4 & 0.665 \\
        CodeLlama 7B Instruct & 31.1\% & 62.2\% & 63.6 & 0.95 \\
        \hline
        DeepSeek Coder 6.7b Instruct & 76.8\% & 79.8\% & 61.2 & 0.38 \\
        DeepSeek Coder 6.7b Instruct & 73.8\% & 88.4\% & 61.1 & 0.665 \\
        DeepSeek Coder 6.7b Instruct & 73.8\% & 90.9\% & 61.1 & 0.95 \\
        \Xhline{1pt} 
    \end{tabular}
    \label{tab:top-p_change}
\end{table}

Table \ref{tab:top-p_change} demonstrates a similar trend when changing \textit{top-p} values. A higher \textit{top-p} setting does not affect the number of tokens generated per second. When inspecting the pass@1 score for CodeLlama, there is a downward trend when increasing the \textit{top-p} value. The DeepSeek Coder model gets a lower pass@1 score when increasing the \textit{top-p}. Both models’ pass@10 scores increase as the parameter moves closer to the baseline.

The results of this experiment align well with our expectations. Both models see an increase in pass@10 when the \textit{temperature} and \textit{top-p} move closer to the baseline, as predicted. When inspecting the results of the pass@1 score, the expectation was for the score to get progressively lower when moving further away from the baseline. The expectation for \textit{top-p} was almost accurate for both models, except for CodeLlama during the \textit{temperature} experiment. In the \textit{temperature} experiment for CodeLlama, we saw a higher pass@1 score when the \textit{temperature} increased from 0.32 to 0.56, which did not align with our expectations. As previously stated, pass@1 tests are more susceptible to randomness and other factors and as such, this can be a result of that. 

Our last expectation stated that the tokens per second should decrease with higher parameter values. The fact that both parameters, designed to loosen constraints and potentially increase output, had no effect is unexpected. Higher \textit{top-p} values give the model more potential tokens to consider, while higher \textit{temperature} widens the probability distribution and thus also elevates the number of viable tokens. Therefore, we believed that increasing parameters would negatively impact throughput because more tokens are considered. However, in practice, these parameters only affect the randomness of the output. One explanation could be that this is a relatively small part of the Large Language Model's (LLM’s) inner workings, and as such, we cannot see any effect on throughput.

These results highlight the sensitivity of model performance to variations in the \textit{temperature} and \textit{top-p} parameters. The findings suggest that specific parameter settings may enhance the model's performance for the pass@10. When inspecting the results for the pass@1 score, we can determine that it affects the results, but how it affects them is unclear.

\Subsection{Prompt Sensitivity Analysis}
\label{sec:prompt_analysis}
These results demonstrate the prompt customization functionality in LLMSecCode which satisfies R2 (Section \ref{sec:key_functionality}).

As described in Section \ref{sec:experiments_plan}, we perform prompt sensitivity analysis by considering four different prompts. The different prompts for the experiment can be found in Listing \ref{lst:prompts}.

Table \ref{tab:prompt_sensitivity_results} presents the prompt sensitivity analysis results for the CodeLlama 7B Instruct and DeepSeek Coder 6.7b. Notably, each prompt yields a unique pass rate. Specifically, prompt 2 achieves the highest pass rate on CodeLlama 7B Instruct, while prompt 4 is most beneficial for DeepSeek Coder 6.7b Instruct. The worst-performing prompt for CodeLlama is prompt 4. For DeepSeek Coder, the worst-performing prompt was number 2. A noteworthy observation is that the best-performing prompt for one model is the worst-performing for the other. The average tokens generated display some patterns. Namely, prompts 3 and 4 generate the most and the least amount of tokens, respectively, for both models.

\begin{table}[htbp]
    \centering
    \caption{Prompt sensitivity results for CodeLlama 7B Instruct \& DeepSeek Coder 6.7b Instruct. The models underwent testing on SecurityEval with one answer for each of the 121 tasks and greedy decoding. Pass rate is the number of non-vulnerable (i.e., no CWEs present) answers divided by total answers.}
    \setlength{\tabcolsep}{5pt} 
    \begin{tabular}{ c | c | c | c }
        \Xhline{1pt} 
        \textbf{Model} & \textbf{Prompt} & \textbf{Pass rate} & \textbf{Tokens per prompt}\\
        \hline
        CodeLlama 7B Instruct & 1 & 50.4\% & 251\\
        CodeLlama 7B Instruct & 2 & 56.2\% & 221\\
        CodeLlama 7B Instruct & 3 & 48.8\% & 300\\
        CodeLlama 7B Instruct & 4 & 47.1\% & 192\\
        \hline
        DeepSeek Coder 6.7b Instruct  & 1 & 45.5\% & 288\\
        DeepSeek Coder 6.7b Instruct  & 2 & 43.8\% & 263\\
        DeepSeek Coder 6.7b Instruct  & 3 & 44.6\% & 318\\
        DeepSeek Coder 6.7b Instruct  & 4 & 49.6\% & 235\\
        \Xhline{1pt} 
    \end{tabular}
    \label{tab:prompt_sensitivity_results}
\end{table}

These results indicate that even minor prompt modifications can significantly impact the outcome, which aligns with our expectations. Each prompt, when altered, yields a distinct pass rate on the SecurityEval dataset. Furthermore, these subtle changes also influence the number of tokens generated per response, emphasizing the importance of prompt customization. 

Another interesting finding from this experiment is that one of the prompts can achieve the best performance for one model and the worst for another. However, contrary to our expectation, prompt 3 did not yield the highest pass rates for any model. Moreover, prompts 1, 2, and 3 did not gradually achieve better results with each increasing number. One explanation might be how each LLM interprets the instructions and generates the answers. For example, prompt 3 generates the most tokens for both models, suggesting that both LLMs write more code or explanations, which could cause difficulties with the code extraction method (Section \ref{sec:extraction}).

\Subsection{Comparison with other evaluations}
These results demonstrate the satisfaction of R3 (Section \ref{sec:methods}) in LLMSecCode.

Table \ref{tab:comp_humaneval} compares our Code Llama Instruct HumanEval results with \cite{code_llama}. Notably, our results are slightly higher across all model variations, with a difference of +3.0\% as the highest for the 7B version.

\begin{table}[H]
    \centering
    \caption{Comparison between our evaluation and \cite{code_llama} on the HumanEval dataset. The pass@1 scores are computed with greedy decoding. We label the pass@1 scores as either \textit{previous} (referring to the results of \cite{code_llama} or \textit{our} (referring to our evaluation results).}
    \setlength{\tabcolsep}{5pt} 
    \begin{tabular}{ c | c | c | c | c | c }
        \Xhline{1pt} 
        \textbf{Model} &
        \textbf{Size} &
        \textbf{Dataset} & 
        \textbf{\makecell{Pass@1 \\ (previous)}} &
        \textbf{\makecell{Pass@1 \\ (our)}} &
        \textbf{Difference}\\
        \hline
        \multirow{3}{*}{Code Llama Instruct} & 7B & \multirow{3}{*}{HumanEval} & 34.8\% & 37.8\% & +3.0 \%\\
        & 13B & & 42.7\% & 43.9\% & +1.2 \%\\
        & 34B & & 48.2\% & 48.2\% & +0.0 \%\\
        \Xhline{1pt}
    \end{tabular}
    \label{tab:comp_humaneval}
\end{table}

Our evaluation results consistently receive higher results than \cite{code_llama} The 7B model shows the most significant improvement, with a +3.0\% increase in pass@1 score. The 13B model also improves by +1.2\%. However, the 34B model performs on par with previous results, showing no difference.

In Table \ref{tab:testing}, we compare our results with those of \cite{bhatt_purple_2023}. The 7B model is excluded from this comparison as it was not tested by \cite{bhatt_purple_2023}.

\begin{table}[H]
    \centering
    \caption{Comparison between our evaluation and \cite{bhatt_purple_2023} in the CyberSecEval instruct test. The pass rates of our models are computed with \textit{temperature} of 0.6 and \textit{top-p} of 0.95.}
    \label{tab:testing}
    \begin{tabular}{ c | c | c | c | c }
        \Xhline{1pt} 
        \textbf{Model} &
        \textbf{Size} &
        \textbf{\makecell{Average \\pass rate \\(previous)}} &
        \textbf{\makecell{Average \\pass rate \\(our)}} &
        \textbf{Difference}\\
        \hline
        \multirow{2}{*}{Code Llama Instruct} & 13B & 64.8\% & 70.2\% & +5.4 \%\\
         & 34B & 62.3\% & 66.6\% & +4.3 \%\\
        \Xhline{1pt}
    \end{tabular}
\end{table}

Our results demonstrate a significant improvement in the average pass rates over those reported by \cite{bhatt_purple_2023} For the 13B model, our average pass rate is 71.7\%, an increase of +7.0\% compared to the 64.7\% pass rate reported previously. Similarly, the 34B model shows an average pass rate of 67.6\%, representing an improvement of +5.3\% over the previous 62.3\%.

These results indicate the differences between our evaluation results in the CyberSecEval instruct test, particularly at the 13B scale, where we observe the most significant gain.

Despite our expectations, LLMSecCode yielded surprising results compared to previous benchmarks. Our use of quantized models alone has yielded strong results, with some models surpassing previous benchmarks. One possible explanation for this unexpected outcome is the various factors at play during the test runs, such as prompt, extraction method, and generation configuration, all of which influence the results. Most of these factors are eliminated for CyberSecEval Instruct since the dataset suite facilitates the prompts and extraction processes, reducing a few factors. Nevertheless, an increased performance is observed despite the use of quantized models. 

One hypothesis is that the method for identifying insecure code does not verify if the response is code or just text, potentially resulting in a higher pass rate when the model produces more text. Contrary to expectations, this hypothesis could clarify why the quantized models achieve higher pass rates for the CyberSecEval Instruct test. It is important to note that the generation parameters may not always align precisely with those used by the authors due to a lack of clarification. Additionally, differences in evaluation metrics and preprocessing techniques between studies may introduce biases or inconsistencies. Due to these variables, results indicate a correct implementation since they align within a margin of error.

\Subsection{Secure coding evaluation}
These results demonstrate the satisfaction of R4 (Section \ref{sec:methods}) in LLMSecCode.

\begin{figure}[htbp]
\centering
\includegraphics[scale=0.60]{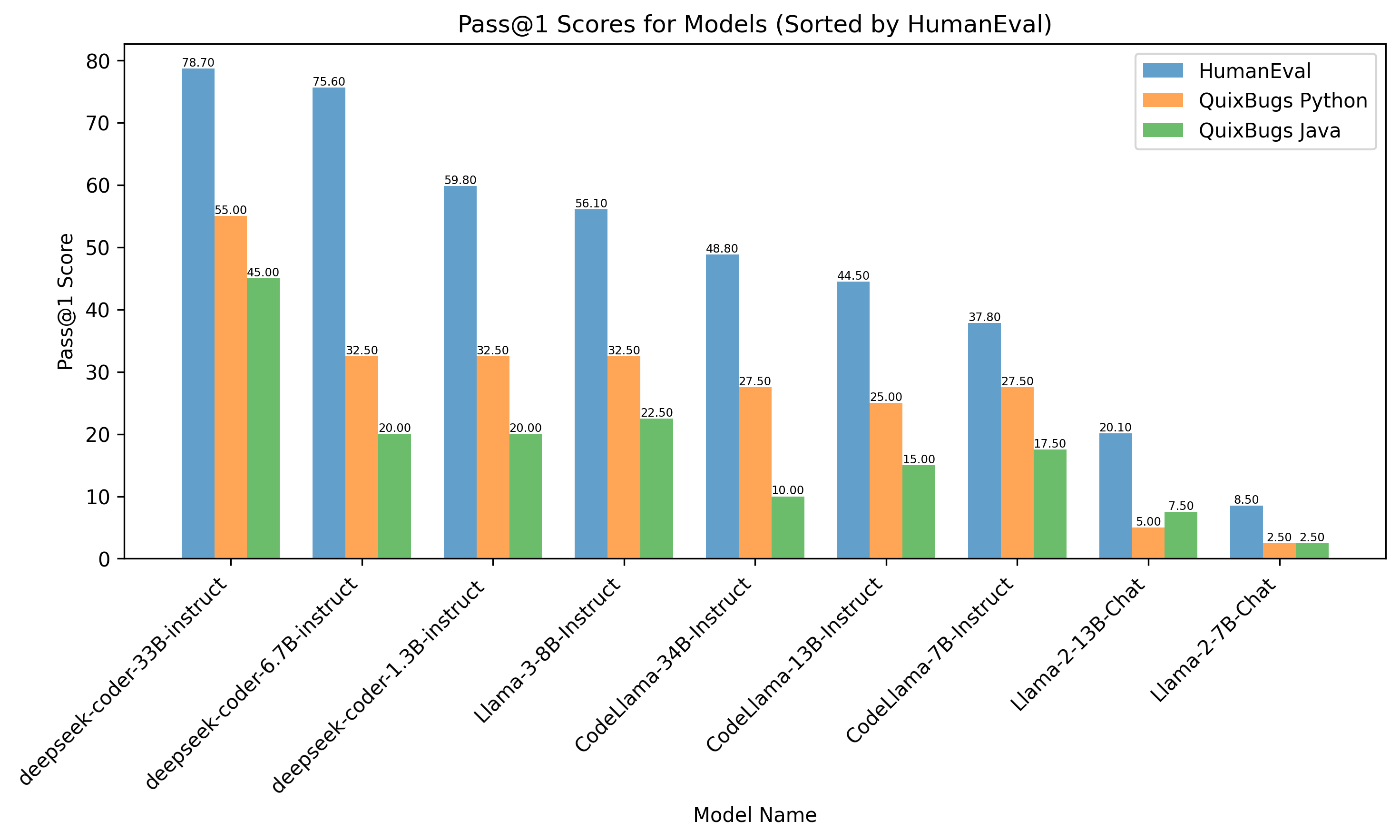}
\caption{HumanEval and QuixBugs results generated with greedy decoding and \textit{max new tokens} of 400 for HumanEval and 1000 for QuixBugs. The figure is sorted by best to worst performance on HumanEval.}
\label{fig:heqb}
\end{figure}

Figure \ref{fig:heqb} shows the baseline created with the QuixBugs and HumanEval datasets. Deepseek Coder 33B Instruct achieves the highest score overall, while Llama 2 7B chat performs the worst. As expected, almost all models trained on code achieve much higher scores than chat-focused LLMs. However, one outlier is the general-purpose Llama 3 model ~\cite{meta_llama_team_introducing_2024}, which outperforms all tested versions of CodeLlama. Furthermore, a trend is that models perform better in Python compared to Java. Another noticeable aspect is that the smallest CodeLlama model (CodeLlama 7B Instruct) performs better than its other versions in Java. Finally, we see that models that perform well on HumanEval also achieve higher scores on QuixBugs Python.

\begin{figure}[htbp]
\centering
\includegraphics[scale=0.60]{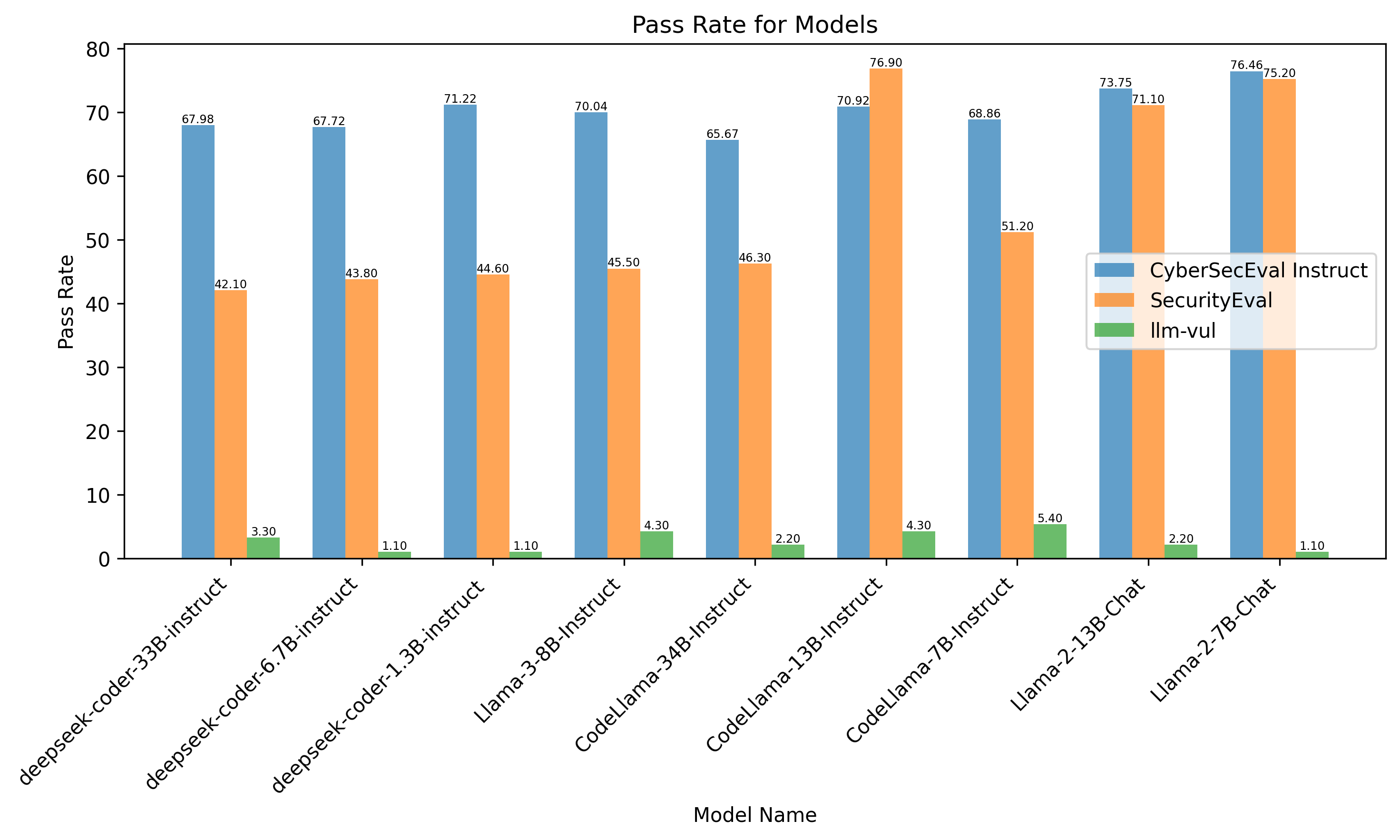}
\caption{CyberSecEval Instruct, SecurityEval, Vul4J and VJBench (llm-vul) results. Generated with greedy decoding and \textit{max new tokens} of 800 for CyberSecEval Instruct, SecurityEval with 400 tokens, and llm-vul with 2000 tokens. The graph has the same order as in Figure \ref{fig:heqb}.}
\label{fig:seceval}
\end{figure}

Figure \ref{fig:seceval} demonstrates the pass rate of models in security-related tasks. Overall, Llama 2 7B performs best, while a general trend is that the better the HumanEval result, the worse the SecurityEval result, except for the outlier CodeLlama 13B. It is also worth noting that the Llama2 models and CodeLlama 13B receive notably higher SecurityEval scores than the rest. Another trend in the results is that the larger versions of each model perform worse than the same model with a smaller size on CyberSecEval Instruct. When inspecting the llm-vul results, there does not seem to be any clear trend. However, as expected, all models achieve low pass rates as the tasks in llm-vul prove challenging.

Our expectations for this experiment were primarily correct. We predicted that higher HumanEval pass@1 scores would lead to lower pass rates in CyberSecEval Instruct and SecurityEval tasks. Furthermore, we found a correlation between HumanEval and QuixBugs Python results. However, we cannot observe a strong relationship between QuixBugs Java and llm-vul which share similar tasks. This might be due to the inherent difficulty of llm-vul tasks, which, in conjunction with finding vulnerabilities, require high \textit{max new tokens} and long coherent responses. As such, no model performs well on these tasks. Some models perform well in solving and creating programs for simple problems. However, the results from CyberSecEval Instruct and SecurityEval suggest that models generate large amounts of code that do not follow standard security practices.

%% file: sections/disucssion.tex
\Section{Discussion}
Each experiment presented in Section \ref{sec:experiments_plan} has the common theme of testing the reliability of the implementation while also showcasing LLMSecCode's capabilities. This section discusses the results and provides some direction for future work.

\Subsection{Discussion of experiment results}
In this section, we discuss each experiment result presented in Section \ref{sec:results}. While each result represents model capabilities in a specific setting, it is worth noting that they do not necessarily reflect the best performance one can achieve in that setting. This is due to many factors, such as the code extraction method, prompts, parameter values, and randomness.
\subsection{Parameter Sensitivity Analysis}
The parameter sensitivity analysis investigated how parameter changes predictably affect the results. The analysis revealed that increasing \textit{temperature} and \textit{top-p} did not significantly impact the number of tokens generated per second. 

Both models' pass@k scores increased in pass@10 for \textit{temperature} and \textit{top-p} values closer to the baseline. As either parameter rises, so does its 'creativity' or its number of viable tokens. This is potentially why the pass@10 scores increase while we generally see a decrease in pass@1. With a higher parameter value and more attempts per task, the model can explore several different, more creative solutions, which could result in a more significant probability of finding a viable answer. 

In contrast, pass@1 scores did not reveal any patterns for CodeLlama or DeepSeek Coder; however, they saw an increase in pass@1 scores with lower \textit{temperature} changes. The small sample size could explain why the CodeLlama model did not have the same expected pattern as the DeepSeek Coder. As such, the randomness can have a more significant influence on the results. Investigating the parameters' effect on pass@1 may require redesigning the testing methodology.

\Subsubsection{Prompt Sensitivity Analysis}
Our experiment highlights that different prompts work best for various models. As such, prompts also affect how prone a model is to generate code that does not follow standard security practices. However, our experiment is not perfect, as Large Language Models (LLMs) are highly abstract. SecurityEval does not check the correctness of each solution. As such, a higher pass rate does not necessarily mean that the respective prompt is best suited for generating code and solving specific problems with security in mind. Therefore, to improve the experiment, one would have to create test suites that check the correctness of each solution and exclude those that do not pass all tests.

As each prompt achieved different results, the output may have some bias. The bias can be due to the training data for each model. Examining the data is necessary to determine the root of such a bias. The training data's wording could influence prompt sensitivity, potentially explaining why certain prompts lead to better results on specific models. Notably, CodeLlama exhibits a slightly better overall performance in this regard, suggesting its potential for generating more secure code.

\Subsubsection{Comparison with other evaluations}
Our comparison experiment resulted in the tested models achieving better results with LLMSecCode compared to previous evaluations. The evaluations that we use for comparison do not disclose all possible details about their experiments. Therefore, exact recreation is challenging. We encourage researchers to include more information about their experiments for more accurate reproduction.

\Subsection{Secure coding evaluation}
Many aspects of our SC evaluation can be experimented with to determine the capabilities of LLMs more accurately. For example, utilizing infilling versions of models to simulate other Automated Program Repair (APR) techniques. This would require pinpointing and communicating the vulnerable lines of code in each APR task prompt, which could reveal that infilling is more appropriate for such problems. Furthermore, as briefly mentioned in Section \ref{sec:prompt_analysis}, creating test suites for SecurityEval and CyberSecEval Instruct to exclude solutions that do not solve the task would yield more comparable outcomes. 

We have intentionally made the tasks of llm-vul difficult to more accurately simulate real-world scenarios. These tasks proved to be challenging, and none of the tested LLMs achieved a high pass rate. An alternative approach to these tasks is to focus the model on all vulnerable lines instead of the entire function. This would most likely yield higher pass rates for all models. Furthermore, it would focus more on the model's vulnerability-fixing capabilities. However, in a real-world setting, these lines would not be apparent.

Another issue with our evaluation is its focused nature. Both Code Generation (CG) and APR tasks test few programming languages. As such, a major improvement to LLMSecCode would be to expand its diversity to include more languages. Moreover, HumanEval and QuixBugs do not test real-world problems but are focused on human-made tasks. This would be another improvement to create a broader baseline in an assessment similar to our SC evaluation. Implementing datasets such as MultiPL-E (\cite{cassano_multipl-e_2022}) and Defects4J (\cite{just_defects4j_2014}) would be a good place to start.

One feature that was not explored in this evaluation is the correction chain (Section \ref{sec:CoT}). This could potentially yield interesting results for both SecurityEval and llm-vul. By providing what CWE and errors that a model's previous solution resulted in, it is possible that it could produce more secure answers. However, this feature would not work on CyberSecEval since LLMSecCode simply extends this suite and does not modify the testing of its benchmarks.

We found that the LLMs performance in security-oriented tasks is insufficient to allow independent development. This might be due to their training data. We speculate that these models are not necessarily trained on SC practices, but rather learn patterns from real-world code that implements it. As such, models might not be aware or understand secure practices at all. Therefore, one suggestion is to test models specifically trained on SC, which might achieve much higher pass rates in security-oriented tasks.


\Subsection{Future work}
Due to limitations in time and resources, several promising avenues for further testing exist. Our SC evaluation is just one path within a broader set of possible evaluation routes. Some interesting tests could be: 
\begin{itemize}
  \item \textbf{Instruction and infilling comparison:} As we allow benchmarks on both conversation types, exploring their performance differences could be interesting. Such an experiment could reveal that one type is better suited for specific tasks.
  \item \textbf{Model security:} One feature of LLMSecCode is the ability to utilize open-source models with the CyberSecEval test suite, enabling the evaluation of many different models on specific benchmarks designed to test model security. Combining these with the other implemented datasets could yield interesting results.
  \item \textbf{Explore correction chain capabilities:} With LLMSecCode, it is possible to allow the model to correct its answer (see Section \ref{sec:CoT}). We did not explore this feature in this report. However, it could let users see how effectively a model corrects itself in different areas and presents an interesting avenue for future work.
\end{itemize}

Another important aspect that we want to emphasize is the LLMSecCode's continued development. Many decisions were made during the creation of LLMSecCode, specifically for developers who want to improve it. As such, we encourage any willing developers to implement more datasets, models, and features. This would ensure relevancy and further improve comprehensiveness. 

One significant concern during this project was the need for appropriate datasets for testing and benchmarking LLMs in APR and CG. To further improve LLMSecCode, there is a pressing need for more datasets that cover a broader range of programming languages and diverse aspects of each area. Additionally, existing datasets often focus on a limited range of coding scenarios, failing to encompass the full complexity of real-world programming tasks. Expanding and enriching these datasets will enhance the robustness and versatility of benchmarks, making them more effective in evaluating LLMs.

Finally, we encourage all LLM developers to experiment with LLMSecCode to potentially enhance their models in various areas. By leveraging this framework, developers can explore new techniques, identify and address weaknesses, and ultimately create more robust, efficient, and accurate language models. Collaboration and innovation within the community are vital to driving progress, and LLMSecCode can be a valuable framework in that pursuit.

%% file: sections/conclusion.tex
\Section{Conclusion}
LLMSecCode is a new open-source framework for evaluating LLMs' SC capabilities. Our findings validate that LLMSecCode operates as intended. Our pilot implementation relies on static analyzers and automated processes for labeling some of the studied datasets since they help us sketch the model's performances.
As an enhancement to our work, we propose to include more labeled datasets to further mitigate the effect of these process automation on the accuracy of the results. 
Our preliminary results demonstrate the satisfaction of our requirements (R1-R4, Section \ref{sec:key_functionality} \ref{sec:methods}). Our experiments reveal the impact of parameters and prompts on the studied models as well as expose scenarios in which the state-of-the-art LLMs generate insecure code. We believe that LLMSecCode can help any security researchers and practitioners who wish to systematically evaluate LLM models on different datasets dealing with secure coding.

\subsection*{Acknowledgments}

This work was partially supported by the Swedish Civil Contingencies Agency (MSB) through the projects RIOT/RICS2.

We extend our heartfelt gratitude to Mohannad Alhanahnah for his invaluable assistance in improving the presentation of this article.